\documentclass[a4paper,12pt]{article}
\pdfoutput=1
\usepackage{xcolor}
\usepackage{enumerate}
\usepackage{framed}
\usepackage{mdframed}
\usepackage{paralist}
 \mdfsetup{%
 linecolor=white,
 backgroundcolor=gray!20,
 }
\usepackage{amsmath}
\usepackage{amsfonts}
\usepackage{mathrsfs}
\usepackage{amssymb}
\usepackage[utf8x]{inputenc}
\usepackage{graphicx, rotating}
\usepackage{epsfig}
\usepackage{latexsym}
\usepackage{graphicx}
\usepackage{color}
\usepackage{amsmath,bm,amssymb}
\usepackage{cite}
\usepackage{slashed}
\usepackage{hyperref}
\usepackage{epstopdf}
\usepackage[font=footnotesize,labelfont=]{caption}
\hypersetup{colorlinks=true, citecolor=bluscuro, linkcolor=black, urlcolor=bluscuro}
\definecolor{rossos}{cmyk}{0,1,1,0.55}
\definecolor{bluscuro}{rgb}{0.15, 0.2, .85}
\definecolor{bluchiaro}{cmyk}{1,.3,0.,0.1}
\definecolor{Gray}{gray}{0.95}

\def\bea{\begin{eqnarray}}
\def\eea{\end{eqnarray}}

\newcommand{\llp}{\left [}
\newcommand{\rrp}{\right ]}
\newcommand{\lp}{\left (}
\newcommand{\rp}{\right )}
 \def\be   {\begin{equation}}   \def\ee   {\end{equation}}
 \def\ba   {\begin{array}}      \def\ea   {\end{array}}

\def\mp{M_\text{\tiny Pl}}

\def\circa#1{\,\raise.3ex\hbox{$#1$\kern-.75em\lower1ex\hbox{$\sim$}}\,}

\newcommand{\beq}{\begin{equation}}
\newcommand{\eeq}{\end{equation}}

\font\tenrsfs=rsfs10 at 12pt
\font\sevenrsfs=rsfs7
\font\fiversfs=rsfs5
\newfam\rsfsfam
\textfont\rsfsfam=\tenrsfs
\scriptfont\rsfsfam=\sevenrsfs
\scriptscriptfont\rsfsfam=\fiversfs
\def\mathscr#1{{\fam\rsfsfam\relax#1}}

\renewcommand\d{{\rm d}}

\def \lta {\mathrel{\vcenter
     {\hbox{$<$}\nointerlineskip\hbox{$\sim$}}}}
\def \gta {\mathrel{\vcenter
     {\hbox{$>$}\nointerlineskip\hbox{$\sim$}}}}

\begin{document}

\thispagestyle{empty}
\vspace{0.1cm}
\begin{center}
{\huge \bf 
On the Cosmological Stability\\[1ex] of the Higgs Instability}  \\[20mm]

{\bf\large Valerio De Luca$^{a}$, Alex Kehagias$^b$,  
 Antonio Riotto$^a$}  \\[5mm]

{\it $^a$ D\'epartement de Physique Th\'eorique and Centre for Astroparticle Physics (CAP),\\[-1mm]
Universit\'e de Gen\`eve, Geneva, Switzerland}\\[0mm]
{\it $^b$ Physics Division, National Technical University of Athens, Athens, 15780, Greece}

\vspace{18mm}
{\large \bf Abstract}
\begin{quote}
The Standard Model Higgs potential becomes unstable at large Higgs field values where its quartic coupling  becomes negative. While the tunneling lifetime of our current  electroweak vacuum is comfortably longer than the age of the universe, quantum fluctuations during inflation might push the Higgs over the barrier, forming patches which might be lethal for our universe. 
We  study the cosmological evolution   of such  regions and find that, at least in the thin wall approximation,  they may be  harmless as they    collapse due to the backreaction of the Higgs itself.
The  presence of the   Standard Model Higgs instability can   provide a novel mechanism to end inflation and to reheat the universe through the evaporation of the  black holes left over by the collapse of the Higgs bubbles. 
The bound on  the Hubble rate during inflation may be therefore relaxed. 

\end{quote}
\end{center}

\vfill
\noindent\line(1,0){188}
{\scriptsize{ \\ E-mail:\texttt{  \href{mailto:valerio.deluca@unige.ch}{valerio.deluca@unige.ch}, \href{mailto:kehagias@central.ntua.gr}{kehagias@central.ntua.gr},
\href{mailto:antonio.riotto@unige.ch}{antonio.riotto@unige.ch}
}}}

\pagebreak\large
\tableofcontents
\newpage\normalsize

\section{Introduction}
It is well-known that the current  measured values of the  Higgs boson and top quark masses lead to the  surprising and intriguing realization that,  in the context of the Standard Model (SM) with no additional physics, our universe situates itself  at the edge between stability and instability of the electroweak vacuum
(for some recent studies, see Refs.  \cite{Buttazzo:2013uya,Degrassi:2012ry,Elias-Miro:2011sqh}). 

The SM  Higgs potential develops an instability well below the Planck scale, at an energy scale around $\Lambda \sim 10^{12}$ GeV\footnote{Such a value is estimated taking into account the new measurement of the top quark mass, obtained by combining the latest Particle Data Group average~\cite{ParticleDataGroup:2020ssz} with the new CMS result~\cite{CMS:2022kcl}, but see also Ref.~\cite{Khoury:2021zao} for a recent analysis.} (for a gauge-invariant definition of the instability scale, see Ref.~\cite{Espinosa:2016nld}), but with a  lifetime which is comfortably and  exceedingly longer than the age of the universe. 

The stability of the electroweak vacuum is however not at all guaranteed in the early universe, when the presence of an unbounded from below direction  with negative energy may represent a threat. For instance, during inflation \cite{Lyth:1998xn} the fluctuations $\delta h$ of the Higgs field, if the latter has  no direct coupling to the inflaton and is minimally coupled to gravity (and therefore  effectively massless during inflation), may push the Higgs field away from our  current electroweak vacuum and above the barrier \cite{Espinosa:2007qp} (see also Refs.~\cite{Herranen:2014cua,Hook:2014uia,Espinosa:2015qea} and, for a recent review, Ref.~\cite{Markkanen:2018pdo}). This happens if roughly
\be
\delta h\sim\frac{H_0}{2\pi}\gta \Lambda,
\ee
where $H_0$ is the inflationary Hubble rate, so that the survival probability for the Higgs to remain in our electroweak vacuum is exponentially suppressed with time, $P_{\text{\tiny surv}}\sim {\rm exp}(-H_0^3 t/32 \Lambda^2)$, and the normalised Higgs field variance, which describes the correlation function within surviving domains, approaches a constant $\langle h^2 \rangle/P_{\text{\tiny surv}} \propto \Lambda^2$~\cite{Espinosa:2007qp}.\footnote{The probability of the Higgs vacuum decay may be enhanced in the presence of primordial black holes with masses  $\lta 10^5 \mp$ (where $\mp^2=1/8\pi G$ is the reduced Planck mass)  \cite{Burda:2015isa, Burda:2015yfa, Tetradis:2016vqb,Canko:2017ebb}.}
During (and after) inflation our universe would then be filled up by  spherical bubbles containing large values of the Higgs field probing the unstable region. It  was concluded in Ref.\cite{Espinosa:2015qea} that such anti-de Sitter patches would be  lethal for our universe: at the end of inflation such regions with negative vacuum energy  density would expand at the speed of light and eventually engulf all space of our universe. This leads to the requirement  that the probability to find an expanding  anti-de Sitter  bubble in our past light-cone must be negligible, 
thus  deriving  a strong upper bound on the Hubble constant during inflation (see also Refs.~\cite{Joti:2017fwe,Franciolini:2018ebs})
\be
H_0\lta 5\cdot 10^{-2}\Lambda.
\ee
The goal of this paper is to revisit the issue of the Higgs instability during inflation. We will show that  the Higgs bubbles, originated  by the fact that the  Higgs fluctuations push the Higgs over  the barrier,  in fact may  collapse due to the backreaction of the dynamics of the Higgs itself in the thin wall limit, an effect which has not been previously accounted for in the literature. Our findings indicate that the presence of the SM Higgs instability may not represent a danger for our universe. On the contrary, we will show that it offers a novel way to end inflation, to reheat our universe and to start the standard radiation-dominated phase.

The paper is organised as follows. In Section 2 we discuss the dynamics in the interior of the Higgs bubbles. In Sections 3, 4 and 5 we analyse the fate of the Higgs bubbles and we conclude in Section 6. Three Appendices are devoted to technical details.

\section{The dynamics in the interior of the Higgs  bubbles}
In this Section we are going to discuss the dynamics in the interior of the bubbles which are formed when the Higgs field overcomes the barrier of its potential and  rolls down the potential along the unbounded from below  direction.
Bubble formation can occur either through a stochastic process during inflation (interpretable as a Hawking Moss process~\cite{Linde:1990flp}), or through a Coleman-de Luccia tunneling~\cite{Coleman:1980aw} or through oscillating Fubini instantons~\cite{Lee:2014ula}\footnote{In the following we will use the word “bubble" to indicate the formation of a true vacuum domain, even though technically it refers only to the nucleation mechanism.}.

The  background interior metric describes in full generality  a O$(1,3)$ symmetric Friedmann-Robertson-Walker (FRW) universe and is given by 
\be
\label{mi}
\d s_{-}^2=-\d \eta^2+a^2(\eta)\Big(\d \chi^2+S^2(\chi) \d \Omega_2^2\Big), \qquad S(\chi)=
\begin{cases}
\sin \chi, \\
\chi, \\
\sinh \chi ,
\end{cases}
\ee
where $\d \Omega_2^2$ is the metric on the unit two-sphere and $a(\eta)$ denotes the scale factor in terms of the time coordinate $\eta$ in the interior of the bubble and $S(\chi)$ takes into account the possible different interior geometries.
The equation of motion of the Higgs field is given by
\be
h'' + 3 \frac{a'}{a} h' - \frac{\nabla^2 h}{a^2} + \frac{\partial V_{\text{\tiny eff}}}{\partial h} = 0, \label{h0h}
\ee
where the prime indicates differentiation with respect to $\eta$ and $V_\text{\tiny eff}$ is the effective Higgs potential. The equation of motion for the scale factor reads
\begin{align}
(a')^2&=-k+\frac{ a^2\rho}{3\mp^2},\nonumber\\
\rho&=\frac{1}{2}(h')^2 +\frac{1}{2}\frac{(\nabla h)^2}{a^2} +V_{\text{\tiny eff}},  
\label{fred}
\end{align}
where $k = 0, \pm 1$ identifies a flat, close or open universe. As we will show later, the exact value of $k$ is not important for our results, as a period of inflation will make it irrelevant and the interior of the bubble spatially locally flat.
Since the Higgs values we will be considering probe the instability region, we will assume the quartic coupling to be  negative at large Higgs field values. For  our purposes one can assume the simplified expression for the Higgs potential
\be
V_{\text{\tiny eff}}= V_0 - \frac{\lambda(h)}{4}h^4, \label{vtin}
\ee
in terms of an inflation vacuum energy $V_0$ and the running quartic coupling~\cite{Espinosa:2015qea}
\be
\label{quartic}
\lambda(h)\simeq \frac{0.16}{(4\pi)^2}\ln\left(\frac{h^2}{\Lambda^2\sqrt{e}}\right).
\ee
 To properly solve the system of coupled differential equations, including the spatial dependence of the Higgs field,  we have performed a lattice calculation making use of  the program LATTICEEASY, where the fields and its derivatives are evaluated on a spatial grid with evenly spaced points, in an expanding background~\cite{Felder:2000hq}.
The use of a lattice simulation which fully accounts for non-linear effects  is necessary to properly include the effects of the spatial perturbations of the Higgs field. Indeed, due to the presence of the negative quartic term in the potential, the Higgs perturbations may experience in principle the phenomenon of  tachyonic instability~\cite{Felder:2001kt}. 

Let us call $h_0$ the initial value of the Higgs field at $\eta=0$ when the bubble forms. Typically we expect
$h_0\sim H_0/2\pi$,
where $
H_0^2= V_0/3\mp^2$,
and zero velocity $h'(0)=0$ over a Hubble distance $H_0^{-1}$, but our general results are independent from these assumptions. The Higgs bubble will therefore form with positive energy density inside dominated by the vacuum energy. Negative energy densities would require $h_0^2\gta (H_0\mp/\lambda^{1/2})$, which is highly improbable (for our analytical approaches, in the following we will simplify the Higgs potential by a simpler form  $-\lambda h^4/4$, with $\lambda >0$). Furthermore, the initial values of the Higgs modes are given by quantum fluctuations, where each mode has a random phase and a Gaussian random amplitude with expectation value $\langle |h_k|^2 \rangle = 1/2k$, corresponding to massless fluctuations, on subhorizon scales.

\begin{figure}[t!]
	\centering
	\includegraphics[width=0.53\columnwidth]{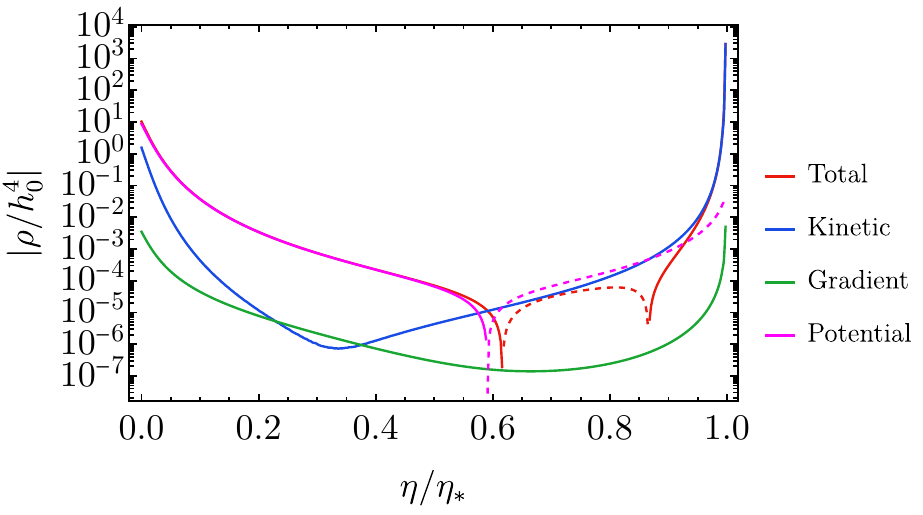}
	\includegraphics[width=0.43\columnwidth]{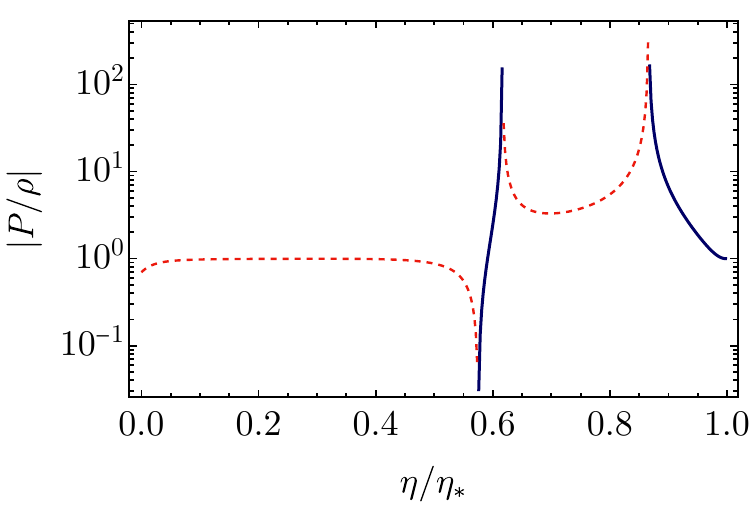}
	\caption{Left: Behaviour of the different components of the energy density with time. We have chosen $H_0/\mp=10^{-4}$, $h_0=H_0$, $h'(0)=0$ and $a_0=1$. Dashed lines indicate a  negative energy density. Right: The corresponding equation of state as a function of time, where $P$ indicates the pressure density.}
	\label{energy}
\end{figure}

In Fig.~\ref{energy} we plot the total energy density of the system and its various components (left) and the corresponding equation of state (right). We highlight  three  points. First of all, the dynamics is characterized by the presence of a singularity point at $\eta=\eta_*$, whose origin  we will discuss later on. Secondly, 
the contribution from the Higgs gradient energy density (computed through  LATTICEEASY by averaging over distances larger than the scale $1/(\sqrt{\lambda} h_0)$, obtained as a combination of parameters of the Higgs potential and initial conditions to normalise momentum scales and to perform spatial averages~\cite{Felder:2000hq}) is always subdominant  compared to the Higgs kinetic and potential 
energy densities. Thirdly, while at the beginning the energy density is dominated by the inflation vacuum energy, the energy density becomes first negative,  when the Higgs potential energy dominates due to the negative quartic coupling, and subsequently positive   when the energy density of the system is dominated by the kinetic energy density of the Higgs field. Notice also that, at this stage, the potential energy density grows much more slowly than the kinetic energy density.

Since the fully non-linear solution of the dynamics shows that  the gradient energy density of the Higgs is negligible, we are allowed to  describe the dynamics of the scale factor and the Higgs field in the interior of the bubble and to explain the origin of the   singularity point without including such a contribution.

We are dealing therefore with a typical potential with an  unbounded from below direction. The equations of motion reduce to
\begin{eqnarray}
&&h''+3\frac{a'}{a}h'+\frac{\partial V_{\text{\tiny eff}}}{\partial h}=0,\nonumber \\ 
&&(a')^2=-k+\frac{ a^2}{3\mp^2}\left(\frac{1}{2}(h')^2+V_{\text{\tiny eff}}\right), \label{eq2}
\end{eqnarray}
while the  equation for the conservation of the energy density reads
\begin{eqnarray}
\frac{\d\rho }{\d \eta}=\frac{\d}{\d \eta}\left(\frac{1}{2}(h')^2+V_{\text{\tiny eff}}\right)=-3 \frac{a'}{a}(h')^2. \label{rhodot}
\end{eqnarray}
The total energy density  is increasing for 
contracting ($a'<0$) or decreasing for expanding ($a'>0$) universe. 

After the formation of the Higgs bubble, that occurs when the Higgs has jumped over the potential barrier, its interior still  experiences a stage of inflation, with the scale factor growing exponentially. The Higgs motion is classical if in a Hubble time the motion is dominated by the classical friction rather than by the quantum one.
If so, the Higgs scalar field slowly rolls down from the top of the effective potential satisfying 
the equation 
\begin{eqnarray}
3H_0 h'\simeq \lambda h^3,
\label{fr}
\end{eqnarray}
which describes the initial motion of the Higgs field.
The classical motion beats the quantum one if \cite{Espinosa:2017sgp}
\be
\delta h_{\text{\tiny c}}\sim \frac{h'}{H_0}\sim  \frac{\lambda h_0^3}{3 H_0^2}\gta \delta h_{\text{\tiny q}}\sim\frac{H_0}{2\pi},
\ee
or
\be
h_0^3\gta \frac{3H_0^3}{2\pi\lambda},
\ee
which we will assume from now on in order for the slow-roll approximation to hold at the initial stages.
The solution  of Eq. (\ref{fr}) is given by
\be
\label{a}
h^2(\eta)\simeq \frac{h_0^2}{1-2\lambda h_0^2 \eta/3H_0}.
\ee
The quartic term starts dominating over the vacuum energy (and the total energy becomes negative, before getting positive again), $\lambda h^4\sim \mp^2 H_0^2$, 
around the singularity  time 
\be
\eta_*\simeq \frac{3H_0}{2\lambda h_0^2}.
\ee
The Higgs zero mode value  at this stage is such that $h^2\sim (\mp H_0/\lambda^{1/2})\ll \mp^2$. 
Subsequently, the speed of the Higgs field rapidly increases and in a Hubble time reaches the singularity. To understand this point, let us neglect for a moment  the  effects of the expansion of the universe in the Higgs equation of motion such that,  at large values of the Higgs field, we may write \cite{Felder:2002jk}
\be
(h')^2\simeq 2(V_0-V_{\text{\tiny eff}}),
\ee
and the acceleration of the universe is dictated by the equation
\be
\frac{a''}{a}=\frac{1 }{3\mp^2}\left[V_{\text{\tiny eff}}-(h')^2\right]\simeq \frac{1 }{3\mp^2}\left(3V_{\text{\tiny eff}}-2V_0\right)\simeq -\frac{\lambda }{4\mp^2}h^4.
\ee
At large Higgs values the universe starts moving with ever growing negative acceleration. If one reintroduces back the expansion of the universe, the Higgs speed becomes even smaller, and the deceleration is even greater. As a result, the expansion slows down and the universe starts contracting. At this stage the friction term  $3(a'/a) h'$ in the equation of motion of the Higgs  becomes negative, which causes the Higgs to grow and leads to a rapid collapse of the universe~\cite{Felder:2002jk}. Therefore, once the universe approaches the turning point where the total energy density vanishes ($a'=0$),  it begins collapsing, and the total energy density becomes positive again.
During the last phase, the  energy density is therefore  dominated by the Higgs kinetic term 
and we may write
\begin{figure}[t!]
	\centering
	\includegraphics[width=0.49\columnwidth]{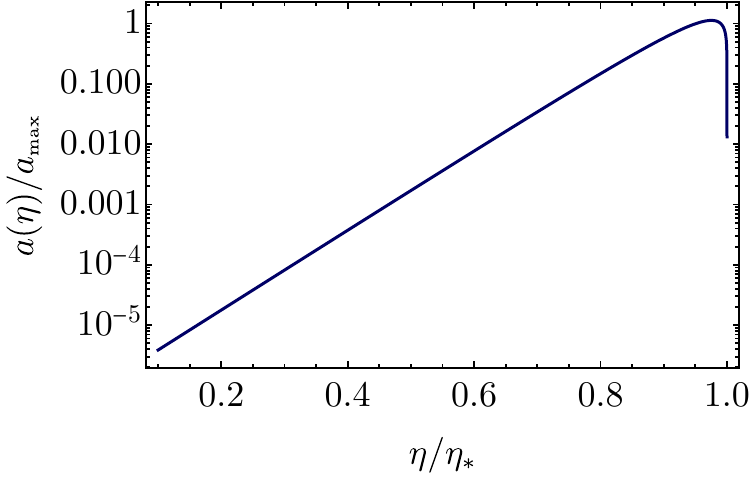}
	\includegraphics[width=0.49\columnwidth]{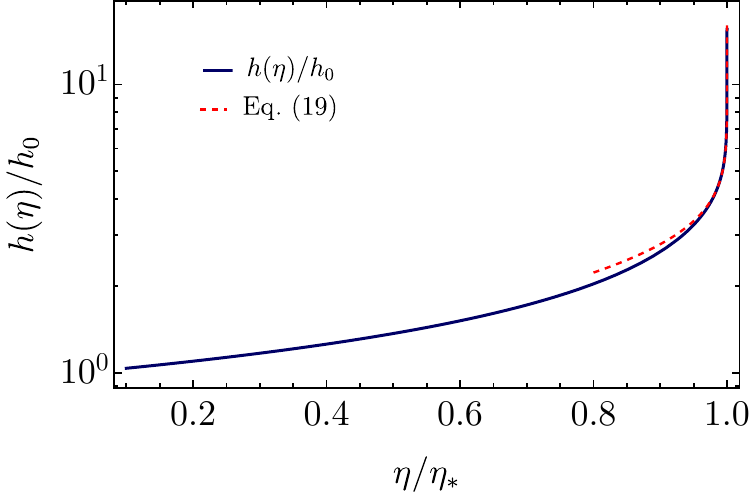}
	\caption{Left: time behaviour of the scale factor close to the singularity, normalised with respect to its maximum value $a_\text{\tiny max}$. Right: time behaviour of the Higgs zero mode close to the singularity. We have chosen $H_0/\mp=10^{-4}$, $h_0=H_0$, $h'(0)=0$ and $a_0=1$.}
	\label{sfh}
\end{figure}
\begin{align}
(a')^2&\simeq \frac{a^2 }{3\mp^2}\frac{1}{2}(h')^2,\nonumber\\
\frac{h''}{h'}&=-3\frac{a'}{a}, 
\end{align}
whose solutions are
\begin{align}
a(\eta)&\simeq a_0(\eta_*-\eta)^{1/3}, \\
h(\eta)&\simeq h_0+ \mp\sqrt{\frac{2}{3}} \ln \left( \frac{\eta_*-\eta_0}{\eta_*-\eta}\right). 
\end{align}
 These behaviours are reproduced by the  numerical results plotted in Fig. \ref{sfh} where we show the evolution of the scale factor and the Higgs  field. Notice also that the potential energy density grows only as $\ln^4 (\eta_*-\eta)$, that is much smaller than the kinetic energy density which scales likes $1/(\eta_*-\eta)^2$ around the pole. This explains the hierarchy between the two contributions in Fig. \ref{energy}.

The curvature term is always  negligible because of the previous inflationary stage. The universe is therefore effectively flat at this epoch. Furthermore, the Higgs field close to the singularity  rolls down slowly, only logarithmically. We assume that, in the case in which there is a new deeper minimum at some high value of the Higgs field, say at  scales larger than the Planck mass (possibly generated by a higher-dimensional operator of the form $h^6/\mp^2$), the scale factor contracts before it is reached. This assumption is needed in case the bubble forms once the field has tunneled very close to the true minimum, even though this scenario is much less probable than diffusing past the maximum of the potential, where our picture would apply, for large Hubble rate.

We are now in the position to go back to the issue of the Higgs gradient energy and explain analytically why it is smaller than the other components. During the period in which the scale factor expands exponentially, the gradients of the Higgs field are suppressed and so it is their contribution to the Higgs total energy density compared to the other components when the Higgs starts rolling down its potential. The equation of motion of the rescaled Higgs field $\widetilde h_k=a h_k$ in Fourier space and in terms of the comoving time $\d\widetilde \eta=\d\eta/a$ reads at the linear level
\be
 \frac{\d^2 \widetilde{h}_k}{\d\widetilde{\eta}^2}+\left(k^2-\frac{1}{a}\frac{\d^2 a}{\d\widetilde{\eta}^2}-3\lambda h^2(\widetilde\eta)a^2\right)\widetilde h_k=0.
\ee
By appropriately choosing the initial condition we can set     $\widetilde\eta\sim (\eta_*-\eta)^{2/3}$ close to the singularity. As the scale factor contracts, the tachyonic mass squared gets smaller and smaller, and modes which are  outside the comoving Hubble radius $\widetilde{{\cal H}}^{-1}=a/(\d a/\d\widetilde{\eta})\sim \widetilde\eta$ have an amplitude $\widetilde h_k\sim a$. During the contraction phase more modes exit the Hubble radius, as the physical wavelength $a/k\sim (\eta_*-\eta)^{1/3}$ scales slower than the physical Hubble radius 
$a/a'\sim (\eta_*-\eta)$, and one expects  the 
contribution to the gradient energy density to be smaller than the  kinetic energy density. To estimate the former we consider  the  Higgs fluctuations which  have an amplitude given roughly by
$h_k\sim H_0/k^{3/2}$~\cite{Espinosa:2017sgp}, as they inherit the de Sitter power spectrum during the slow-roll phase of the Higgs field. Correspondingly, the Higgs gradient energy density scales such that 
\be
\frac{(\nabla h)^2}{2a^2}\sim\int^{ 1/\widetilde \eta}\frac{\d^3 k}{(2\pi)^3} \frac{k^2|h_k|^2}{2a^2}\sim\frac{H_0^2}{\widetilde \eta^2a^2}\sim  \frac{H_0^2}{(\eta_*-\eta)^2}\ll \frac{1}{2}(h')^2\sim \frac{\mp^2}{(\eta_*-\eta)^2}.
\ee
Our numerical results, including the same time dependence of the Higgs gradient  and kinetic energy densities, confirm our  analytical findings.

\section{Setting the stage for the Higgs bubble evolution}
The goal of this Section is to set the stage for the description of the dynamics of the Higgs bubbles once they are formed, in order  to understand if the Higgs bubbles  may or not pose a threat for our current universe.

The Higgs  bubble   materializes as a  spherically symmetric  object. At its interior the metric is given in Eq.~(\ref{mi}), while   the external space  will be that of Schwarzschild-de Sitter with metric 
\begin{eqnarray}
\d s^2_+=-f(r)\d t^2+\frac{\d r^2}{f(r)}+ r^2 \d \Omega_2^2, \qquad f(r)=1-\frac{2GM}{r}-H_0^2 r^2,
\end{eqnarray}
assuming that the initial radius of the bubble is smaller than $1/H_0$ (see Appendix B for the case of initial bubble radii larger than the Hubble radius).
The bubble wall is located at 
\begin{eqnarray}
\chi=X(\tau),
\end{eqnarray}
as seen from the inside geometry and 
at
\begin{eqnarray}
r=R(\tau) 
\end{eqnarray}
from the outside geometry. The intrinsic bubble metric 
is 
\begin{eqnarray}
\d \sigma^2 =\gamma_{ij} \d \xi^i \d \xi^j=-\d \tau^2 +R^2(\tau) \d \Omega_2^2, \label{mb}
\end{eqnarray}
where $\xi^i=(\tau,\theta,\phi)$ are the bubble worldvolume coordinates.
The
 induced metrics  on the bubble from the two sides are  
\begin{eqnarray}
\d \sigma^2_-=-\left(
\dot T_-^2-a^2{\dot X}^2\right)\d \tau^2 +a^2 S^2(X)  \d\Omega_2^2,
\label{m-}
\end{eqnarray}
from the interior 
and 
\begin{eqnarray}
\d \sigma^2_+=-\left(
f(R) \dot T_+^2-\frac{{\dot R}^2}{f(R)}\right)\d \tau^2 + R^2  \d\Omega_2^2, \label{m+}
\end{eqnarray}
from the exterior  and  we have used a dot to indicate derivatives with respect to $\tau$. 
We call the attention of the reader on the fact that times are dubbed as follows
\begin{align}
{\rm interior}:\,\,\,\,\eta&=\eta(\tau)=T_{-}(\tau),\nonumber\\
{\rm exterior}:\,\,\,\,t &= t (\tau)=T_{+}(\tau).
\end{align}
The two metrics (\ref{m-}) and (\ref{m+}) should coincide  with the intrinsic bubble metric of Eq.~\eqref{mb}, leading to the 
conditions
\begin{align}
\dot T_-^2-a^2{\dot X}^2 &=1,
\label{q1}\\
 f(R) \dot T_+^2-\frac{{\dot R}^2}{f(R)}&=1, \label{q2}\\
\frac{R}{aS}&=1.
\label{q3}
\end{align}
We apply the Israel matching conditions \cite{Israel:1966rt}
for the extrinsic curvature of the bubble
\begin{eqnarray}
K^+_{ij}-K^-_{ij}=-\frac{1}{\mp^2} \left(S_{ij}-\frac{1}{2} S_k^k\gamma_{ij}\right), \label{isr}
\end{eqnarray}
where 
\begin{eqnarray}
K_{ij}=e^\mu_{i} e^\nu_j \nabla_\mu n_\nu,
\end{eqnarray}
$n^\mu$ being the normal to the bubble hypersurface, and 
\be
S_{ij}=-2 \sigma \gamma_{ij}
\ee
is the surface stress-energy tensor of the bubble wall in terms of its tension $\sigma$.
The  projection tensor $e^\mu_{i}$ is defined as 
\begin{eqnarray}
e^\mu_{i}=
\frac{\partial x^\mu}{\partial \xi^i}.
\end{eqnarray}
To solve for the dynamics of the Higgs bubble wall analytically, we take 
the thin wall approximation\footnote{This is the same assumption taken  in Ref.  \cite{Espinosa:2015qea} again to  make the problem tractable analytically.},  which assumes that any variation of the Higgs scalar field  along the wall occurs only on length scales much larger than the wall
thickness, that is $\partial_\mu h\sim n_\mu$. It  follows that the bubble tension is constant with time \cite{Blau:1986cw},   a starting point  we will take from now on.
Admittedly, this is a strong assumption and a fully non-linear numerical analysis would be required to go beyond the thin wall approximation.
  
If $u^\mu$ is the velocity vector  along a geodesic, then the normal $n^\mu$ is defined as 
\begin{eqnarray}
u^\mu=\frac{\d x^\mu}{\d \tau}, ~~~~~~u^\mu u_\mu=-1, ~~~~~~
n^\mu u_\mu=0, ~~~~~~~n^\mu n_\mu=1.
\end{eqnarray}
These conditions specify the normal vector, in an obvious notation, as
\begin{eqnarray}
n_-^\mu=\left(a \dot X,\frac{\dot T_-}{a},0,0
\right), ~~~~~~~~~
n_+^\mu=\left(\frac{\dot R}{f},f\dot T_+,0,0\right). 
\end{eqnarray}
Using the above normal vectors, the extrinsic curvature turns out to be
\begin{eqnarray}
K^-_{\theta\theta}= \dot T_- a S \frac{\partial S}{\partial\chi}+\dot X a^2 a'S^2, ~~~~~~~
K^-_{\phi\phi}=\sin^2\theta K^-_{\theta\theta},
\end{eqnarray}
and similarly, 
\begin{eqnarray}
K^+_{\theta\theta}= fR\dot T_+, ~~~~~~~
K^+_{\phi\phi}=\sin^2\theta K^+_{\theta\theta}. 
\end{eqnarray}
We stress that the Israel matching condition for the $\tau\tau$-component is satisfied once the conditions for the other components of the extrinsic curvature are fulfilled, assuming the thin wall approximation~\cite{Blau:1986cw}.
The  Israel matching  condition (\ref{isr}) leads then to the equation
\begin{eqnarray}
fR\dot T_+-\dot T_- a S \frac{\partial S}{\partial\chi}-\dot X a' a^2 S^2=-\frac{\sigma R^2}{2 \mp^2},
\end{eqnarray}
which can be written as 
\begin{equation}
f\dot T_+-\dot T_-  \frac{\partial S}{\partial\chi}-\dot X  a'a S=-\frac{\sigma R}{2 \mp^2}. \label{israel}
\end{equation}
From Eq.~(\ref{q3}) we find that 
\begin{eqnarray}
\dot X=\frac{1}{a^2 \frac{\partial S}{\partial\chi}}\left(-a' \dot T_- R+a\dot R\right),
\label{i1}
\end{eqnarray}
so that Eq.~(\ref{q1}) is expressed  as 
\begin{eqnarray}
 \dot T_-^2-\frac{1}{ {\left(\frac{\partial S}{\partial\chi}\right)^2}}\left(-\frac{a'}{a} \dot T_- R+\dot R\right)^2=1.  \label{i2}
 \end{eqnarray} 
 Solving Eq.~(\ref{i2}) for $\dot T_-$ and using it together with 
 Eq.~(\ref{i1}) in the Israel matching condition Eq.~(\ref{israel}), we get 
 \begin{eqnarray}
\left(\dot{R}^2+1+\frac{R^2}{a^2}-\frac{a'^2}{a^2}R^2\right)^{1/2}
=\epsilon \Big(f+\dot{R}^2\Big)^{1/2}+\frac{\sigma R}{2 \mp^2},
\label{ii}
 \end{eqnarray}
 where $\epsilon={\rm sign}\,{\dot T_+}$. 
 With the help of the Friedmann equation  (\ref{eq2}),  
 we can write Eq. (\ref{ii}) as
 \begin{eqnarray}
  \dot{R}^2+V(R)=-1,  \label{dr}
 \end{eqnarray}
where 
\begin{eqnarray}
V(R,\eta)=- \frac{k_1}{R^4}- \frac{k_2}{R}- k_3 R^2  \label{VR}
\end{eqnarray}
is a time-dependent potential 
with 
\begin{eqnarray}
k_1=\frac{1}{16\pi^2}\frac{M^2}{\sigma^2}, \qquad k_2=\frac{M}{6\pi}\frac{(\rho_{\text{\tiny c}}-\rho)}{\sigma^2},
\qquad  
k_3=\frac{\rho}{3 \mp^2}+\frac{(\rho-\rho_{\text{\tiny c}})^2}{9\sigma^2}.
\label{k123}
\end{eqnarray}
We have defined the constant $\rho_{\text{\tiny c}}$ as
\begin{eqnarray}
\rho_{\text{\tiny c}}= \frac{3\sigma^2}{4 \mp^2}+3 \mp^2 H_0^2. \label{rhoc}
\end{eqnarray}
From Eq. (\ref{dr}) one can write the expression for the mass $M$ as 
\begin{eqnarray}
\label{mass}
 M=\frac{4}{3}\pi R^3(\rho-\rho_{\text{\tiny c}})+
 4 \pi \sigma R^2 \left(1-\frac{\rho}{3 \mp^2} R^2+\dot R^2\right)^{1/2}, \label{MM}
 \end{eqnarray} 
 which is a constant of motion whose sign is not a priori defined. However, 
 negative masses, once the bubble collapses, would leave behind a naked singularity, which is in contradiction with the Cosmic Censorship theorem \cite{Penrose:1969pc}, for more details see Appendix A. Notice also that, as we will discuss later on, neither the magnitude nor the sign of the mass are relevant for the dynamics close to the singularity. 
 
Summarizing, in order to determine the dynamics of the bubble, one should solve the following equations
\begin{align}
f(R) \dot T_+^2-\frac{{\dot R}^2}{f(R)}&=1, \label{q21}\\
 \dot T_-^2-\frac{1}{ \left(\frac{\partial S}{\partial\chi}\right)^2}\left(-\frac{a'}{a} \dot T_- R+\dot R\right)^2&=1,  \label{i22}\\
\dot{R}^2+V(R)&=-1. 
\label{ii2}
\end{align}

\section{The fate of the Higgs bubbles}
After setting the stage to study their dynamics, we are now in the position to investigate the fate of the Higgs bubbles. We do it first numerically and then offer a more analytical insight in the next Section. It is convenient to regard $R$ and $T_+$ as functions of $\eta=T_-$. 
In this case, Eqs. (\ref{q21}-\ref{ii2})   are written as
\begin{align}
{R'}^2&=-(1+V)\left(1-\frac{1}{1+\frac{R^2}{a^2}}\left(-\frac{a'}{a}R+R'\right)^2\right) , \label{eq00}\\
{T_+'}^2&=\frac{1}{f}\left(1+\frac{{R'}^2}{f}\left(1 -\frac{1}{1+\frac{R^2}{a^2}}\left(-\frac{a'}{a}R+R'\right)^2\right)^{-1} \right), \label{eq01}
\end{align}
  and 
  \begin{eqnarray}
  \dot{T}_-^2=\left(1-\frac{1}{1+\frac{R^2}{a^2}}\left(-\frac{a'}{a}R+R'\right)^2\right)^{-1} \label{eq03}. ~  \end{eqnarray}
 Notice that these set of equations exactly reproduce Eqs. (\ref{aa1}-\ref{aa3}) in Appendix B for the case of initial bubble radii larger than the Hubble radius by simply setting $f=1$.
 Solving for $R'$ through Eq. (\ref{eq00}), we find the most relevant  equation of the paper
\be
  R'=-\frac{a a'(1+V)R}{R^2-a^2 V}\pm 
  \frac{1}{R^2-a^2 V}\sqrt{(1+V)(a^2+R^2)\Big(a^2 V-R^2(1-{a'}^2)\Big)}. 
  \label{eq04}
  \ee
 Now the strategy is simple.  Eq. (\ref{eq04}) is a differential equation  which provides   $R(\eta)$.  Then the solution can be used in Eq. (\ref{eq01}) to determine 
  $T_+$. Finally, Eq. (\ref{eq03}) determines $\eta(\tau)$ which can be used to specify the solutions as 
  $R\big(\eta(\tau)\big)$ and $t\big(\eta(\tau)\big)$.
 
\begin{figure}[t!]
	\centering
	\includegraphics[width=0.7\columnwidth]{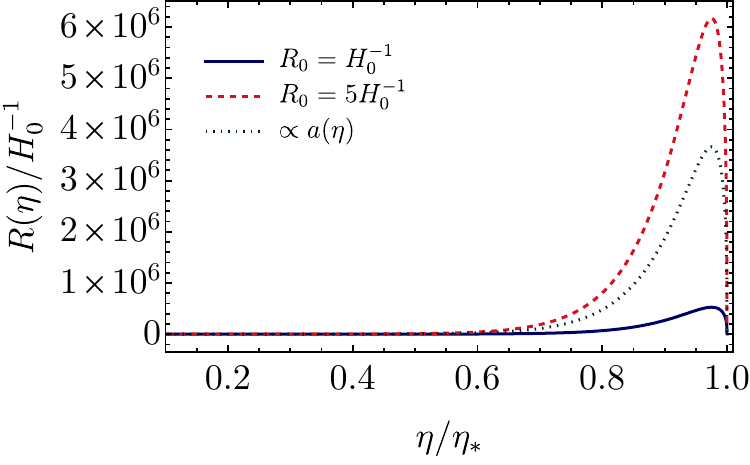}
	\includegraphics[width=0.45\columnwidth]{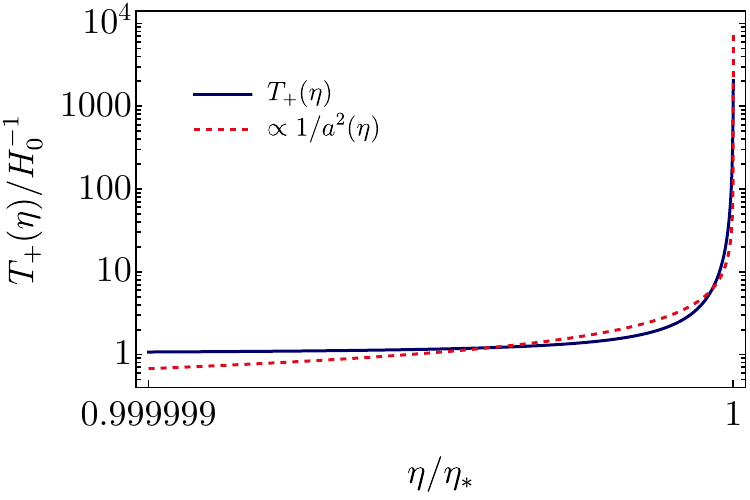}
	\includegraphics[width=0.45\columnwidth]{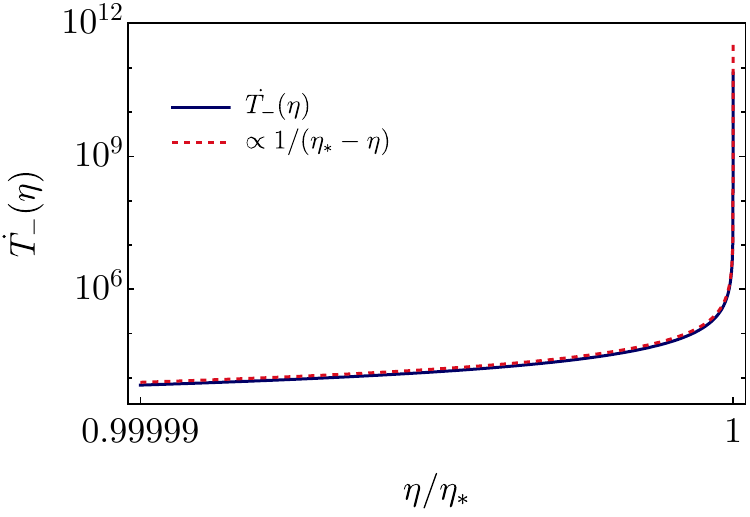}
	\caption{The evolution of the bubble wall radius (top), and external and internal times (bottom). We have chosen $H_0/\mp=10^{-4}$, $h_0=H_0$, $h'(0)=0$ and  we have renormalised the initial value of the scale factor to unity at the beginning of the Higgs dynamics. }
	\label{Radius}
\end{figure}

 The behaviour of the bubble radius is shown in Fig.~\ref{Radius} for different initial conditions. 
 Our results indicate that the Higgs bubbles expand exponentially at the beginning, but end up  contracting and they do so by following the   same time dependence of the internal scale factor. In particular,  the bubbles start increasing their sizes, have a turning point when $R'=0$,   and 
at the  final stage they collapse to zero size, leaving a singularity behind. Eq. (\ref{eq00}) shows that the bubble wall has  a turning point not only when $(1+V)$  vanishes, but also when the second parenthesis does so. Indeed, as we will see later on, at the turning point $V$ is very large in absolute value and $R'$ approaches zero because of the growing behaviour of $\dot{T}_-$.
Our results are not sensitive to  the bubble mass $M$  and to the choice of the initial bubble radius in units of $H_0^{-1}$. 

 \begin{figure}[t!]
	\centering
	\includegraphics[width=0.6\columnwidth]{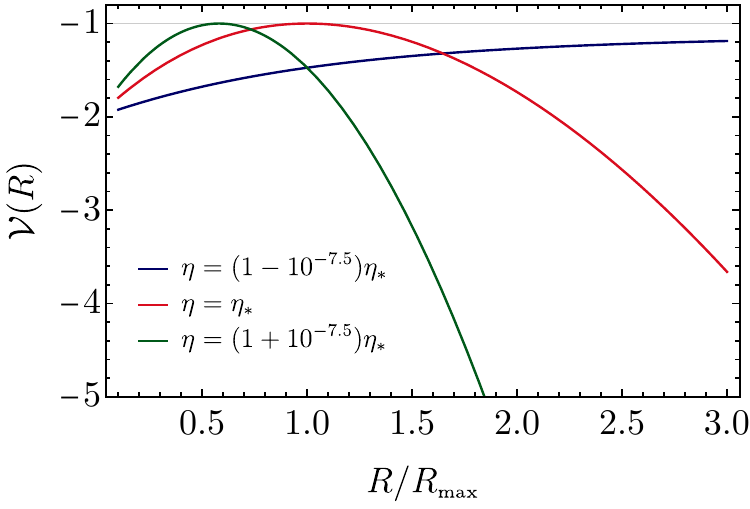}
	\caption{Illustrative plot of the bubble potential $\mathcal{V}(R)$ in terms of the bubble radius $R$ at different times $\eta$ close to the turning point.}
	\label{potential}
\end{figure}
To understand better the behaviour of the bubble wall, let us rewrite Eq.~\eqref{eq04} in the following form
\be
R'^2 + \mathcal{V} (R) = -1,
\ee
by defining the bubble potential
\be
\mathcal{V} (R) \equiv - \llp -\frac{a a'(1+V)R}{R^2-a^2 V}\pm 
  \frac{1}{R^2-a^2 V}\sqrt{(1+V)(a^2+R^2)\Big(a^2 V-R^2(1-{a'}^2)\Big)} \rrp^2 -1.
\ee
For a fixed radial coordinate $R$, the bubble potential depends on time both through the scale factor $a$ and the potential $V$. To show its time dependence we have to consider therefore the time evolution inside the bubble. 
At initial times, the radius of the bubble can be taken of the order of the inverse Hubble horizon, that is $R_0\sim H_0^{-1}$, and the Higgs fluctuations are of the order of $h_0 \sim H_0/2\pi$.  Our results do not depend however from this choice. The wall tension and mass are then given by~\cite{Espinosa:2015qea}
\begin{eqnarray}
\sigma \sim \frac{\sqrt{\lambda}}{(2\pi)^3} H^3_0 \qquad  {\rm and} \qquad 
M \sim \frac{\sqrt{\lambda}}{(2\pi)^2} H_0,
\end{eqnarray}
and, as already pointed out above, they do not depend on time.
At the initial stage the typical parameters which enter in the potential $V$ are then  
\begin{align}
k_1&=\frac{1}{16\pi^2}\frac{M^2}{\sigma^2} \sim  \frac{1}{4 H_0^4}, \nonumber\\
k_2&=\frac{M}{6\pi}\frac{(\rho_{\text{\tiny c}}-\rho)}{\sigma^2} \sim  \frac{\sqrt{\lambda}}{6 \pi H_0},\nonumber\\
k_3&=\frac{\rho}{3 \mp^2}+\frac{(\rho-\rho_{\text{\tiny c}})^2}{9\sigma^2} \sim H_0^2 - \frac{\lambda}{(2 \pi)^4} \frac{H_0^4}{\mp^2} + \frac{\lambda H_0^2}{9(2 \pi)^2} \sim H_0^2,
\end{align}
and one can appreciate the hierarchy
\begin{eqnarray}
\frac{2GM}{R_0}\sim\frac{\sqrt{\lambda}}{(2\pi)^3}\frac{H_0^2}{\mp^2}\ll
H_0^2 R_0^2.
\end{eqnarray}
Once the equations of motion for the scale factor and scalar field have been solved, one can plug the solutions in the effective potential and show its behaviour in terms of the bubble radius, for different times.
The result is shown in Fig.~\ref{potential}.
The crucial information one gets is that, close to the big crunch $\eta_*$, the effective potential crosses $-1$. At that stage the bubble velocity $R'$ becomes zero, and thus the bubble radius starts contracting from its maximum value $R_\text{\tiny max}$  following the behaviour already shown in Fig.~\ref{Radius}. Furthermore, after reaching the turning point, the potential gets squeezed to smaller radii.
Our  results can be understood from the Newtonian point of view. In the limit of $(\sigma/\mp)^2\ll \rho$ and $\dot R\simeq 0$, the bubble mass (\ref{MM})
receives two  contributions, one from the volume term and the other from the surface term. Being both positive, the bubble contracts.

We stress that our findings  do not change by assuming different initial conditions. In particular,   close to the pole the value of the bubble mass $M$   is not relevant for the behaviour of the bubble radius, thus  changing only slightly the value of the radius at the turning point $R_\text{\tiny max}$. We devote the next Section to describe these behaviours analytically.

 \section{Analytical insights}
 To understand the previous results analytically, let us  analyse the dynamics using Eq.~(\ref{eq04}). During the phase when the vacuum energy dominates and the internal scale factor grows exponentially, $a(\eta)\sim {\rm exp}(H_0\eta)$, the bubble wall expands exponentially as well, since Eq.~(\ref{eq04}) reduces to $R'/R\simeq a'/a$. Let us now see what happens when 
  the big crunch for the scale factor  at $\eta=\eta_*$ is approached for typical initial conditions.
As we have previously proven, close to the collapse, the kinetic energy of the Higgs field is dominating and we have 
\begin{eqnarray}
 a\sim (\eta_*-\eta)^{1/3}.  \label{a1}
 \end{eqnarray} 
From Eq. (\ref{eq2}) we find that the energy density scales like 
\begin{eqnarray}
\rho \sim\frac{1}{a^6}\sim \frac{1}{(\eta_*-\eta)^2}. \label{r1}
\end{eqnarray}
Eq. (\ref{eq04}) 
simplifies considerably as one takes the limit of large $V$. Indeed, inspecting Eq.~(\ref{k123}) we see that close to the singularity
\be
\frac{k_1}{R^4}\sim \frac{M^2}{\sigma^2}\frac{1}{R^4}\sim \frac{1}{a^4}\ll \frac{k_2}{R}\sim
\frac{M\rho}{\sigma^2}\frac{1}{R}\sim 
\frac{1}{a^7}\ll k_3R^2\sim 
\frac{\rho^2}{\sigma^2}R^2\sim \frac{1}{a^{10}},
\ee
so that
\be
V\sim -\rho^2R^2\sim -\frac{R^2}{(\eta_*-\eta)^4}
\ee
and the  mass $M$ of the bubble does not play any crucial role in the dynamics. Eq. (\ref{eq04}) reduces  to 
\begin{eqnarray}
R'\approx \frac{a'}{a}R\pm \frac{\sqrt{a^2+R^2}}{-a},  \label{eq100}
\end{eqnarray}
which, inserted in Eq.~(\ref{eq00}),  shows that the bubble wall velocity becomes zero as the second parenthesis vanishes. Eq.~(\ref{eq100}) can be solved at leading order close to the singularity as  
\begin{eqnarray}
R(\eta) \sim a(\eta)\sim (\eta_*-\eta)^{1/3}. \label{r12}
\end{eqnarray}
The full numerical result also confirms the analytical expectation that, close to the singularity, the bubble size and the scale factor have the same time dependence, see Fig.~\ref{Radius}. 
Plugging this solution into the equations for $T_+$ and $T_-$ one  easily finds that,  
close to the singularity,
\be
{T_+'}^2\sim -V\sim \frac{R^2}{(\eta_*-\eta)^4},
\ee
leading to the external observer time 
\be
t= T_+(\eta)\sim \frac{1}{a^2}\sim\frac{1}{(\eta_*-\eta)^{2/3}}.
\ee
The external observer sees the bubble collapsing with a time behaviour 
\be
R(t)\sim t^{-1/2}\,\,\,\,{\rm for}\,\,\,\,t\rightarrow\infty.
\ee
Similarly, we find 
\be
{\dot T_-}^2\sim -\frac{V}{R'^2}\sim \frac{1}{(\eta_*-\eta)^2}\,\,\,\,{\rm or}\,\,\,\,
\tau\sim (\eta_*-\eta)^2\sim t^{-3},
\ee
in such a way that the bubble contracts for the proper  observer on the bubble as $R(\tau)\sim \tau^{1/6}$ around the singularity $\tau= 0$.

The  difference between our results and those found in Ref.~\cite{Espinosa:2015qea} is that there the potential $V(R)$ was considered to be  time-independent, since it was assumed that the Higgs field was already sitting at a deep minimum  at some value $h\gta \mp$, thus neglecting the time-dependent dynamics of the Higgs. In the interior of the bubble the geometry was exactly anti-de Sitter and in the exterior exactly de Sitter or Minkowski. In such a case, the backreaction of the dynamics of the Higgs,  which makes  the total energy density within the bubble positive and large  close to the singularity, is absent and solutions with expanding bubble walls can be found, see Appendix C for a more detailed discussion. This can be understood also in the Newtonian limit where the gravitational self-energy of the wall can be neglected. Large bubbles  grow indefinitely because the system gains energy in the process if the interior energy density is negative,  as in Ref. \cite{Espinosa:2015qea}. In our case,  bubbles  shrink for opposite energetic reasons when the energy density becomes positive due to the Higgs dynamics. Notice also that, during the contraction phase, the vacuum energy of the inflaton field does not play any role. We therefore expect that Higgs bubbles which are formed at the very end of inflation will contract even if they are surrounded by an asymptotic   Minkowski spacetime immediately after inflation.

\subsection{Beyond the thin wall limit}
\noindent
Our considerations are valid in the thin wall limit and going beyond it would require a full numerical approach to include the spatial and time dependence of the Higgs configuration,  taking into account the time dependence of the bubble wall thickness,  the surface energy density, and the surface tension. 
However, it has been suggested that, in the case of a thick bubble, even though the details of the collapse may change (for instance the effective wall tension and the bubble contraction velocity), the results could be qualitatively similar to the thin wall ones~\cite{Carone:1989nj, Garfinkle:1989mv, Barrabes:1993cn,Khosravi:2006dt, Freivogel:2007fx}.

This might be  confirmed by the following arguments. If the Higgs bubble wall thickness is non-vanishing, one has to impose the Israel matching conditions  for the extrinsic curvature of the bubble also on the $\tau\tau$-component. However, using the covariant conservation of the energy-momentum tensor, this turns out to be the time derivative of the $\theta\theta$-component~\cite{Barrabes:1993cn,Khosravi:2006dt} and our system of equations can be used upon   redefining an effective  time dependent  bubble wall tension. If the latter 
decreases in time, our  analytical insights  indicate that the collapse would persist, as the leading term in the potential $V$ close to the singularity would  scale like  $V \sim k_3 R^2 \sim \rho^2 R^2/\sigma^2$, even faster than $\sim \rho^2$. Consider now the more  likely case that  the Higgs bubble wall tension increases. This may happen by (boldly)  assuming  a finite and time-independent thickness $\Delta R$ of the Higgs bubble wall~\cite{Barrabes:1993cn,Khosravi:2006dt}, such that the tension  would roughly scale with time as
\be
\sigma\sim \rho\, \Delta R \sim \frac{\Delta R}{a^6}.
\label{thick}
\ee  
The  various coefficients in the potential $V$ would then have the  behaviour close to the singularity
\begin{align}
k_1&=\frac{1}{16\pi^2}\frac{M^2}{\sigma^2} \sim 0, \nonumber\\
k_2&=\frac{M}{6\pi}\frac{(\rho_{\text{\tiny c}}-\rho)}{\sigma^2} \sim \frac{M}{\mp^2} ,\nonumber\\
k_3&=\frac{\rho}{3 \mp^2}+\frac{(\rho-\rho_{\text{\tiny c}})^2}{9\sigma^2} \sim \frac{\rho}{\mp^2},
\end{align}
and one immediately deduces that the leading term in the potential $V$ close to the singularity is $V \sim k_3 R^2 \sim \rho R^2/\mp^2$, 
leading again to Eq.~(\ref{eq100}) and to a contracting behaviour of the bubble radius, $R \sim a$. We have checked numerically that the bubble radius indeed contracts also in this case.
 Notice that Eq. (\ref{thick}) is  consistent  with the  perturbative expansion at first-order  in the bubble thickness $\Delta R$~\cite{Barrabes:1993cn}
\be
\sigma \sim \Delta R \lp \frac{\dot R^2}{R^2} + \frac{1}{R^2} \rp \sim \frac{\Delta R}{\tau^2} \sim   \rho\,\Delta R \sim \frac{\Delta R}{a^6},
\ee
where the third passage is valid since, in such a case, the   time variables are related to each other by the
scaling $\tau \sim (\eta_* - \eta)$. As stressed above, the resolution of the full problem requires the study of the  spatial and time dependence of the Higgs configuration\footnote{A full general relativity simulation has been performed in Ref.~\cite{East:2016anr} for the case of a bounded Higgs potential with a minimum at large energies, finding that the bubble wall can decouple from a contracting bubble interior and expand, engulfing the universe.}.
If the bubble wall is not thin, we expect the Higgs configuration to be divided in three regions. In the  internal contracting one the Higgs has large values and its positive  kinetic energy density dominates. In the external region the Higgs  vacuum expectation value is zero. In the middle region the Higgs interpolates between the two values.  If the middle  value goes down rapidly enough to continuously follow the internal value such that its energy density becomes positive, we expect the Higgs bubble wall to contract. In the thin wall limit the middle and interior regions coincide, and the Higgs bubble wall contracts following the collapse of the interior region.

\section{Implications and conclusions}
Assuming that the SM holds up to large energies, we have investigated the cosmological evolution of the Higgs field during inflation and reanalysed the issue of the stability of the  electroweak vacuum.  We have argued that the stability may be  guaranteed thanks to the backreaction of the Higgs dynamics. The essential point is that, at the interior of the Higgs patches where the Higgs potential is unbounded from below, the Higgs kinetic energy quickly dominates and the bubble walls collapse  following the  internal scale factor  towards the  singularity. 

This result, which is independent from the particular form of the Higgs potential at large values of the Higgs field in the unbounded from below region,  is intimately connected with the fact that a period of cosmological inflation is incompatible with a universe dominated by a negative energy density. Indeed, after a period of inflation, the curvature term may be omitted and the Friedmann equation may not have a solution with negative energy density. In other words,  the  prediction of the  inflationary cosmology is that we cannot  end up in a space dominated by negative energy density \cite{Linde:2001ae}. When the  interior of the bubble  approaches the turning point, where the total energy density vanishes, it has to collapse in such a way that the  total energy density becomes positive again.
At the same time, the bubble wall follows the internal scale factor due to the  dynamics of the Higgs field itself and collapses. Notice also that the  null energy condition assures  that, if the universe switches from expansion to contraction, it will not  return to the regime of expansion later on, even including the backreaction of particle production.

The singularities left over by the collapse of the Higgs bubbles are hidden  behind a black hole horizon according to the Cosmic Censorship theorem.  
This implies the following dynamics. Inflation starts and the Higgs fluctuations push the Higgs zero mode beyond the barrier after a fraction of one e-fold if $H_0\gg \Lambda$. Subsequently inflation keeps going on for a number of e-folds of the order of $\eta_*H_0$, which is typically much larger than the minimum number of e-folds $\sim 60$ required to reproduce our observable flat universe. Afterwards, all Hubble regions of initial volume $\sim H_0^{-3}$ contained in the inflated region of volume $\sim H_0^{-3} {\rm exp}(3\eta_* H_0) $ collapse under the backreaction of the Higgs field which takes over the energy density, thus stopping inflation. Black holes form and  populate the universe. Their typical mass is of the order of the Hubble rate during inflation and therefore   rapidly evaporate with a time scale of the order of $\sim M^3/\mp^4\sim H_0^3/\mp^4$.
Thus,  possibly and interestingly,  the presence of the SM Higgs instability may offer a novel way to end inflation and to   reheat the universe through minuscule primordial black holes if they evaporate completely (see also Ref. \cite{Kearney:2015vba}). It will be interesting to investigate the consequences of such new dynamics.

Our findings indicate that the Hubble rate during inflation may be  not bounded from above to be smaller than a fraction of the instability energy of the SM Higgs potential. This has implications for the possible future  detection of primordial gravitational waves generated during inflation as their detection requires values of the Hubble rate much larger than the Higgs instability scale~\cite{AlvesBatista:2021gzc}. 

Finally, let us stress again that our results are based  on some assumptions. First, as already remarked,  the thin wall limit which allows a manageable analytical computation. It will be interesting and pressing to  extend our results relaxing the thin wall hypothesis to understand if a thick wall can decouple the expansion of the Higgs bubble from the collapse of its inner region. 
Secondly, the hypothesis that the Higgs rolls down along the unbounded from below direction for a sufficiently large time to allow the Higgs kinetic energy to dominate and the Higgs bubble to collapse. We plan to investigate these issues in the next future.

\subsubsection*{Acknowledgments}
We thank A. Strumia and N. Tetradis for interesting comments and discussions,
G. Felder for insightful comments on the program LATTICEEASY,
and R. Bravo, S. Ellis, J.R. Espinosa, G. Franciolini, G.F. Giudice, D. Racco, and L. Senatore for useful feedbacks.
V.DL. and  A.R. are supported by the Swiss National Science Foundation (SNSF), project {\sl The Non-Gaussian Universe and Cosmological Symmetries}, project number: 200020-178787. A.K. is supported by the PEVE-2020 NTUA programme for basic research  with  project number 65228100.

\vskip 1cm

\appendix

\renewcommand{\theequation}{A.\arabic{equation}}

\setcounter{equation}{0}
\section{The big crunch singularity and the Higgs
dynamics}
We  describe here the fate of a Higgs bubble and show that  it ends up accommodating a singularity in its interior. We are inspired by  Ref.~\cite{Abbott:1985kr}, where the interested reader can find additional details.
Let us consider a scalar field coupled to gravity with  dynamics described by the action
\begin{eqnarray}
S=\int \d^4 x \sqrt{-g}\left\{\frac{\mp^2}{2}R-\frac{1}{2}(\partial h)^2  -V_{\text{\tiny eff}}(h)\right\}.
\label{ac1}
\end{eqnarray}
$V_{\text{\tiny eff}}(h)$ is the effective potential  which we assume to have the form of Eq.~\eqref{vtin}.
We are interested in the era when the negative quartic term in $V_{\text{\tiny eff}}(h)$ dominates the vacuum energy $V_0$, or in other words, when $V_{\text{\tiny eff}}<0.$ During this period, the bubble interior can be approximated by an FRW open universe. 
Then, the background metric has the form of Eq.~(\ref{mi}) and using the equations of motion Eqs. (\ref{h0h}) and (\ref{fred}), we find that
\begin{eqnarray}
a''= -\frac{a}{3\mp^2} \left[(h')^2-V_{\text{\tiny eff}}\right]<0,
\end{eqnarray} 
which implies that the scale factor is concave. Hence, the scale factor increases at the beginning, it reaches a maximum and then it  decreases. Similarly,   multiplying Eq. (\ref{h0h}) by  $h'$ we get Eq. (\ref{rhodot})
and, therefore, the energy $\rho=(h')^2/2+V_{\text{\tiny eff}}$ is increasing in the 
contracting ($a'<0$)  or decreasing in the   expanding ($a'>0$) phase.
At this point one may ask if a singularity forms during the evolution of the universe. This happens if the conditions of Penrose's theorem are  satisfied. The latter ensures that a spacetime cannot be null geodesically complete if the following conditions are fulfilled:
\vspace*{-2mm}
\begin{enumerate}
\setlength\itemsep{-0.25em}
    \item[1)]  the spacetime has a closed trapped surface,
  \item[2)]  the spacetime has a non-compact Cauchy hypersurface,
    \item[3)]  the spacetime Ricci tensor satisfies $R_{\mu \nu} k^\mu k^\nu \geq 0$ for all null vectors $k^\mu$.
\end{enumerate}
Let us start with condition 1) by considering null vectors $\ell^\mu$ and $n^\mu$ such that
\begin{eqnarray}
\ell^\mu\ell_\mu=0, ~~~\ell^\mu\nabla_\mu \ell^\nu=0, ~~~
n^\mu n_\mu=0,~~~\ell^\mu n_\mu=-1 .
\end{eqnarray}
We define then the expansions $\theta_\ell$ and $\theta_n$ as
\begin{eqnarray}
\theta_\ell=\Big(g^{\mu\nu}+\ell^\mu n^\nu+\ell^\nu n^\mu\Big) \nabla_\mu \ell_\nu, ~~~~~
\theta_n=\Big(g^{\mu\nu}+\ell^\mu n^\nu+\ell^\nu n^\mu\Big) \nabla_\mu n_\nu. 
\end{eqnarray}
A trapped surface~\cite{Penrose:1964wq,Hawking:1973uf} exists if the  expansions $\theta_\ell$ and $\theta_n$  for (future-directed) out-going $\ell^\mu$ and in-going $n^\mu$ null geodesic congruences, respectively, satisfy~\cite{Booth:2005qc}
\begin{eqnarray}
\theta_\ell<0\,\,\,{\rm and}\,\,\, \theta_n<0.
\label{theta}
\end{eqnarray}
For a spacetime described by a  metric
\begin{eqnarray}
\d s^2=-\d \eta^2+A^2(\eta,\chi)\d \chi^2+B^2(\eta,\chi)\d \Omega_2^2,
\end{eqnarray}
the conditions (\ref{theta}) turn out to be
\begin{eqnarray}
\pm \frac{1}{A}\frac{\partial B}{\partial\chi}+ B'<0. 
\end{eqnarray}
For the metric in Eq.~\eqref{mi}, this condition translates into
\begin{eqnarray}
a'\pm \coth \chi<0. 
\end{eqnarray}
Then, since $\coth \chi\geq 1$, we find that trapped surface is always formed when~\cite{Abbott:1985kr}
\begin{eqnarray}
a'< -1.
\label{trapcond}
\end{eqnarray}
In our case, the evolution of the universe driven by the scalar field $h$ is the following:
at the initial time $\eta=0$, a bubble is created, which then starts expanding. During its expansion, the scalar field energy density decreases. At a certain point the scale factor reaches a maximum at the moment when $a'=0$, after which the universe starts contracting. During  the contraction phase, the energy density increases unboundedly 
as we approach $a=0$. This implies that $a'$ is always decreasing, and at a certain point it will turn less than $-1$, satisfying  the apparent horizon condition of Eq.~\eqref{trapcond} for sufficiently large $\chi$. Therefore, an apparent horizon will be formed and a trapped surface will always exist.

Condition 2) is also satisfied as it is known that a spacetime with metric of the form of Eq. (\ref{mi}) is globally hyperbolic \cite{Beem:1996xpa} and therefore it has a (non-compact) Cauchy hypersurface. 

Finally, from the Einstein equations one can easily show that, for any null vector $k^\mu$, the Ricci tensor satisfies the inequality
\begin{eqnarray}
R_{\mu\nu}k^\mu k^\nu = \frac{1}{\mp^2}\Big(k^\mu \partial_\mu h \Big)^2\geq 0,
\end{eqnarray}
satisfying the last condition 3).
Therefore,  all conditions of the Penrose's theorem are fulfilled  and the considered spacetime is geodesically incomplete. This implies that the spacetime described by the metric (\ref{mi}) will sooner or later end up in a big crunch singularity during the Higgs dynamics.

\renewcommand{\theequation}{B.\arabic{equation}}

\setcounter{equation}{0}

\section{Initial radius outside the de Sitter Hubble radius}
The analysis of Section 3 applies rigorously only to the case in which the initial radius of the bubble is  smaller than $1/H_0$. In the case in which the bubbles form with an initial size larger than the Hubble radius, the  external metric can be taken to be
\begin{eqnarray}
\d s^2_+=-\d t^2+e^{2H_0 t}\Big(\d \rho^2+ \rho^2 \d \Omega_2^2\Big), \qquad r>1/H_0.  
\end{eqnarray}
This time the  bubble is located at 
\begin{eqnarray}
\chi=X(\tau),
\end{eqnarray}
as seen from the inside geometry and 
at
\begin{eqnarray}
\rho=r(\tau) 
\end{eqnarray}
from the outside geometry. 
The
 induced metrics  on the bubble from the two sides are  
\begin{eqnarray}
\d \sigma^2_-=-\left(
\dot T_-^2-a^2{\dot X}^2\right)\d \tau^2 +a^2 S^2(X)  \d\Omega_2^2,
\label{mm1-}
\end{eqnarray}
and 
\begin{eqnarray}
\d \sigma^2_+=-\left(
 \dot T_+^2-e^{2H_0 T_+}\dot r^2\right)\d \tau^2 + e^{2H_0 T_+}r^2  \d\Omega_2^2. \label{mm1+}
\end{eqnarray}
The two metrics (\ref{mm1-}) and (\ref{mm1+}) should coincide  with the intrinsic bubble metric of Eq.~(\ref{mb}). This leads to the 
conditions
\begin{equation}
\dot T_-^2-a^2{\dot X}^2 =1,
\label{q11}
\end{equation}
\begin{equation}  \dot T_+^2-e^{2H_0 T_+}\dot r^2=1, \label{q12}
\end{equation}
\begin{equation}
R=a S= e^{H_0T_+} r.
\label{q13}
\end{equation}
We obtain
\begin{eqnarray}
n_-^\mu=\left( a \dot X,\frac{T_-}{a},0,0
\right), ~~~~~~~~~
n_+^\mu=\left(e^{H_0 T_+} \dot r,\frac{\dot T_+}{e^{H_0 T_+}},0,0\right), 
\end{eqnarray}
and
\begin{eqnarray}
K^+_{\theta\theta}= e^{H_0 T_+} r\dot T_+ + H_0r^2e^{3H_0 T_+} \dot r, ~~~~~~~
K^+_{\phi\phi}=\sin^2\theta K^+_{\theta\theta}. 
\end{eqnarray}
Correspondingly, the Israel matching  turns out to be 
\begin{eqnarray}
e^{H_0 T_+} r\dot T_++H_0r^2e^{3H_0 T_+} \dot r-\dot T_- a S \frac{\partial S}{\partial\chi}-\dot X a' a^2 S^2=-\frac{\sigma R^2}{2 \mp^2}, 
\end{eqnarray}
which can be written as 
\begin{equation}
\dot T_++H_0re^{2H_0 T_+} \dot r-\dot T_-  \frac{\partial S}{\partial\chi}-\dot X  a'a S=-\frac{\sigma R}{2 \mp^2}. \label{israel2}
\end{equation}
From Eq. (\ref{q13}) we find that 
\begin{align}
\dot X&=\frac{1}{a^2 \left(\frac{\partial S}{\partial\chi}\right)}\left(-a' \dot T_- R+a\dot R\right), \nonumber \\
\dot r&=e^{-H_0 T_+}\Big(\dot R-H_0 R \dot T_+\Big),
\label{i11}
\end{align}
so that Eqs. (\ref{q11}) and (\ref{q12})  are written now as 
\begin{align}
\dot T_-^2-\frac{1}{ \left(\frac{\partial S}{\partial\chi}\right)^2}\left(-\frac{a'}{a} \dot T_- R+\dot R\right)^2&=1,  \label{aa1} \\
\dot T_+^2-\Big(\dot R-H_0 R \dot T_+\Big)^2&=1. 
 \label{aa2}
 \end{align} 
 Solving Eqs.~(\ref{aa1}) and (\ref{aa2}) for $\dot T_-$ and $\dot T_+$, and using  Eq. (\ref{i11})   in the Israel matching condition  (\ref{israel2}), we finally get 
 \begin{eqnarray}
\left(\dot{R}^2+1+\frac{R^2}{a^2}-\frac{a'^2}{a^2}R^2\right)^{1/2}
=\epsilon \Big(1-H_0^2 R^2+\dot{R}^2\Big)^{1/2}+\frac{\sigma R}{2 \mp^2}.
\label{aa3}
 \end{eqnarray}
 We notice that 
for $R>1/H_0$ the system of Eqs.~(\ref{aa1}-\ref{aa3}) reproduces Eqs.~(\ref{q21}-\ref{ii2}) as for large radii the term proportional to $M/R$ can be neglected. Clearly, if the bubble collapses for a de Sitter space, it will also collapse for a Schwarzschild-de Sitter space. The reason is that, if the bubble collapses against the repulsive  de Sitter force $\sim H^2 R$, will do the same when the collapse is assisted  by an attractive $GM/R^2$ force when the bubble mass is positive.

\renewcommand{\theequation}{C.\arabic{equation}}

\setcounter{equation}{0}

\section{Comparison with the past literature}
In this Appendix we wish to compare our results with those found in Ref. \cite{Espinosa:2015qea}. There 
it was assumed that in  the interior of the bubble the  geometry was exactly anti-de Sitter with a cosmological constant given by $-V_{\text{\tiny in}}$, corresponding to the length scale  $1/\ell_{\text{\tiny in}}^2=(V_{\text{\tiny in}}/3\mp^2)$,  and in the exterior with a cosmological constant 
given by $V_{\text{\tiny out}}$, corresponding to the length scale  $1/\ell_{\text{\tiny out}}^2=(V_{\text{\tiny out}}/3\mp^2)$.
It  was found there that bubbles of anti-de Sitter expand in a Minkowski spacetime if  $1/\ell_{\text{\tiny in}}^2>(\sigma^2/4\mp^4)$ for two cases: either  the initial radius is large, or  it is  small but the initial velocity is large enough. In the first case, independently from the bubble mass $M$, one has at large radii
\begin{eqnarray}
\dot{R}^2+V=-1\,\,\,\,{\rm and}\,\,\,\,V\simeq-k_3 R^2,
\end{eqnarray} 
 which can be integrated immediately to give
\begin{eqnarray}
R(\tau)\sim\frac{1}{\sqrt{k_3}} \cosh\Big(\sqrt{k_3} \tau\Big).  
\label{rrr}
\end{eqnarray}
In the second case we have
\begin{eqnarray}
\dot{R}^2+V=-1 \,\,\,\,{\rm and}\,\,\,\,V\simeq-\frac{k_1}{R^4},
\label{ii202}
\end{eqnarray} 
 whose solution is 
\begin{eqnarray}
R(\tau)\sim k_1^{1/6}\tau^{1/3}.  
\label{rrrr}
\end{eqnarray}
Let us show how to reproduce these results starting from the basic Eq.~(\ref{eq04}).   
Let us first consider the case of large radii, for which one can neglect the bubble mass, and reproduce Eq.~(\ref{rrr}).  One obtains\footnote{Notice that AdS is not singular. Comparing to Eq.~(\ref{trapcond}), one  has $a'=\cos(\eta/\ell_{\text{\tiny in}})\geq -1$ and therefore it does not have a closed trapped surface. The singularity is only a coordinate one.}
 \begin{eqnarray}
 a(\eta)= \ell_{\text{\tiny in}} \sin\frac{\eta}{\ell_{\text{\tiny in}}}, \qquad \rho=- \frac{3\mp^2}{\ell_{\text{\tiny in}}^2}, \qquad \rho_{\text{\tiny c}}= \frac{3}{4}\frac{\sigma^2}{\mp^2}+\frac{\mp^2}{ \ell_{\text{\tiny out}}^2}, \qquad k_1=k_2\simeq 0,
 \end{eqnarray}
 \begin{align}
k_3&=-\frac{1}{\ell_{\text{\tiny in}}^2}+\frac{1}{9\sigma^2}\bigg(3\mp^2\left(\frac{1}{\ell_{\text{\tiny out}}^2}+\frac{1}{\ell_{\text{\tiny in}}^2}\right)+\frac{3}{4}\frac{\sigma^2}{\mp^2}\bigg)^2\nonumber \\
&=+\frac{1}{\ell_{\text{\tiny out}}^2}+\frac{1}{9\sigma^2}\bigg(3\mp^2\left(\frac{1}{\ell_{\text{\tiny out}}^2}+\frac{1}{\ell_{\text{\tiny in}}^2}\right)-\frac{3}{4}\frac{\sigma^2}{\mp^2}\bigg)^2\nonumber\\
&\simeq\frac{\mp^6}{V_{\text{\tiny in}}}>0.
 \end{align}
 Close to the coordinate singularity (which we set at  $\eta=0$) one can approximate 
 $a(\eta)\sim \eta$, and assuming    $R\gta 1/\sqrt{k_3}$, Eq. (\ref{eq04}) is written 
 \begin{eqnarray}
 R'\approx \left(k_3 R-\frac{1}{R}\pm\sqrt{k_3(k_3R^2-1)}\right)\,  \eta,  
 \end{eqnarray}
 which goes to zero close to the turning point at $R=1/\sqrt{k_3}$. Expanding around the turning point, we find that the solution is approximately
 \begin{eqnarray}
 R(\eta)\approx\frac{1}{\sqrt{k_3}}\left(1+\frac{k_3^2}{8}\eta^4\right),
 \label{rst}
 \end{eqnarray}
 showing that the bubble radius expands.
 We  can now solve directly the equations of motion for the bubble, which for the present case are written as
 \begin{eqnarray}
  \dot T_-^2-\frac{1}{ 1+\frac{R^2}{T_-^2}}\left(-\frac{1}{T_-} \dot T_- R+\dot R\right)^2=1.  \label{i2201}
\end{eqnarray} 
 Using the above expression for $R(\tau)$ in Eq. (\ref{i2201}) we find that 
 \begin{eqnarray}
 \dot T_-= \pm \frac{\cosh \left(\sqrt{k_3}\,  \tau \right) \llp \sqrt{2} k_3 T_-^2 + \sqrt{2} \cosh^2 \left(\sqrt{k_3}\,  \tau \right) \rrp}{\sqrt{k_3} T_- \sqrt{1+2 k_3 T_-^2 + \cosh \left(2\sqrt{k_3}\,  \tau \right)}} - \frac{\sinh \left(2\sqrt{k_3}\,  \tau \right)}{2\sqrt{k_3} T_-}.
 \end{eqnarray}
We cannot integrate the above equation,  but we can find the solution $\eta=T_-(\tau)$ close to $\eta=0$,  which turns out to be
 \begin{eqnarray}
\tau=\pm\frac{\sqrt{k_3}}{2}T_-^2.
 \end{eqnarray}
 Therefore, the solution for the wall radius close to $\eta=0$ is written as 
 \begin{eqnarray}
 R\approx \frac{1}{\sqrt{k_3}}\left(1+\frac{k_3}{2}\tau^2\right)=\frac{1}{\sqrt{k_3}}\left(1+\frac{k_3^2}{8}\eta^4\right),
 \end{eqnarray}
 which coincides with Eq. (\ref{rst}). 

For small radii, Eq. (\ref{eq04}) reduces
to (selecting the $+$ sign)
\be
R'\simeq 2\frac{k_1}{R^5}\eta,
\ee
which is solved by
\be
R(\eta)\sim k_1^{1/6}\eta^{1/3}\sim \ell_{\text{\tiny in}}^{2/3}\eta^{1/3}.
\ee
The corresponding equation for $T_-$ reads
\be
\dot T_-= \frac{-(k_1/\tau)^{1/3} + \sqrt{[9 + (k_1/\tau^4)^{1/3}][T_-^2 + (k_1 \tau^2)^{1/3}]}}{3 T_-},
\ee
which can be solved close to $\eta = 0$ to find
\be
T_- = \sqrt{\frac{3}{2}}\tau,
\ee
reproducing therefore the behaviour found in Eq.~\eqref{rrrr}. The maximum of the barrier is at $R_{\text{\tiny max}}\sim \ell_{\text{\tiny in}}/(\ell_{\text{\tiny in}}\mp)^{2/3}$, which is reached within  a time $\tau\sim \ell_{\text{\tiny in}}/(\ell_{\text{\tiny in}}\mp)^2\ll \ell_{\text{\tiny in}}$, signalling that the bubble wall passes the maximum of the potential and expands. 

The basic difference among the results of Ref. \cite{Espinosa:2015qea} and the present paper is due to the time dependence of the potential $V$: when the dynamics of the Higgs is included, it changes the dynamics of the bubble wall, creating   a sort of attractor solution which obliges the bubble wall to evolve as the internal scale factor, thus leading to its collapse before the end of inflation.

\bibliographystyle{JHEP}
\bibliography{draft}

\end{document}